\documentclass{article}      % Specifies the document class
\usepackage{latexsym}
\usepackage{epsfig}
\usepackage{amssymb, amsmath}

\begin{document}             % End of preamble and beginning of text.

\title{Application of Thermodynamics to the Reduction of Data
Generated by a Non-Standard System}  % Declares the document's title.
\author{David Ford}      % Declares the author's name.
\date{Department of Physics, Naval Postgraduate School, Monterey Ca.}
\maketitle

%\organization{Department of Physics, Naval Postgraduate School, Monterey Ca.}
%\netaddress{dkf0rd@netscape.net}

%\date{}      % Deleting this command produces today's date.

%\maketitle                   % Produces the title.

\section{Introduction}        % Produces section heading.  Lower-level
                             % sections are begun with similar 
                             % \subsection and \subsubsection commands.

The purpose of this paper is to illustrate the application of of thermodynamics as a
pedagogy for the organization, dimensional reduction and analysis of data. To 
accomplish this it is necessary to build a bridge between the world of everyday
experience(a sequence of events occurring in time) and the more abstract world
of constant energy surfaces, microcanonical distributions and the notions of
work and heat. 

Both worlds meet at the path, although an experimentalist may speak of it in
terms of
observations contained in a data record and a theorist may speak of trajectories
in a phase space. In order to not obscure the core concepts, more abstract orderings
or partial orderings of primitive events are not discussed but the careful reader
will be able to find appropriate launch points off the main storyline for 
application specific digressions.

The initial connections are made using the Birkhoff version of the Ergodic Theorem
but before implications of that theorem are spun off some
preparatory review of equilibrium thermodynamics and necessary conventions for reference
frames are introduced. 

Further fundamental connections are made once the Principle of Matched Invariants is introduced.
This principle, and its implications, nail down the temperature field in time up to an ambiguity in the 
sense of orientation and a choice of scale.  The orientation ambiguity is resolved 
insisting on the notion of an absolute zero and an appeal to casuality.
The notion of heat and work are introduced along with the first
law of thermodynamics. 

Warning: The second law is not introduced as a foundational principle but instead,
as an example of a nonstandard application of the foregoing first law apparatus
(and as a way to enter a discussion of the 
second law in a concrete setting) the thermodynamic 
interpretation of a simple closed queueing model is presented. 
It is argued that the context of that discussion allows for the least upper
bound of the heating rate to be specifically identified.

\section{ Definitions and Conventions}

In classical equilibrium thermodynamics it is supposed that the trajectory of the total
system moves under the influence of its Hamiltonian and that at all times the trajectory
lies on a surface of contant energy, i.e. that total system energy is conserved.

Of special interest in this work is the situation where only a subset of the total system
is observed directly. Energies resident in the modes of the observed subsystem at a particular 
time are considered
to be fluctuations derived from the total system as it moves on the constant energy surface.

In this case, the total system Hamiltonian is considered to be the sum of a Hamiltonian
which governs the motion of the bath (the total system minus the subsytem) and a 
second Hamiltonian which governs the motion of the subsystem itself.
Concerns about the lack of an interaction Hamiltonian are not relevant
to the present purpose.

In the event that the energy of the total system is changed, the trajectories of the total system leave the 
former surface (at the old constant energy) and inhabit a new surface(at the new total energy).
These energies surfaces are indexed by temperature. In what follows the raising or lowering
of the total system trajectory from one energy surface to another is taken to be, by convention,
a temperature change.

Mathematically, the same effect of raising or lowering the surface on which the total system
energy resides could be achieved  say, by addition of a  
reference energy to the subsystem Hamiltonian. In this work, by convention,
once a reference energy is selected further uniform change of the subsystem
state energies is interpreted as a temperature change of the total system.

From the point of view of the subsystem, an arbitrary reassignment of state energies may always be uniquely decomposed
into  changes about the old
reference point which leave it unchanged (herein termed zero-sum changes), followed 
by a uniform shift in the reference point(interpreted as a temperature change in the bath).
The zero sum changes will be interpreted as changes to the subsystem Hamiltonian. 

For the sake of simplicity, consider a finite system with finitely many possible energies.\footnote{
As an assurance for the reader please note that the conventions adopted in this section 
leave us with subsystems and subsystem Hamiltonians the properties of which are consistent with and 
are in fact shared by the popular Ising Model
for a bounded portion of a lattice magnet in finite dimensions. Further, a projection
of the total dynamics onto N states may be accomplished in a variety of ways. Presentation
of such a computation, at this point, would obscure the development of the fundamentals.}
Let the set of states available to the subsystem be denoted by

\begin{equation}\label{upsilon}
\boldsymbol{\upsilon} = \lbrace \upsilon(1), \upsilon(2),\ldots ,\upsilon(N)\rbrace           
\end{equation} 

A $\it{data \; record}$ consists of a finitely long sequence of states visited along with the 
length of time each visit in the sequence lasted. For any given data record there
is a rare state. That is, the one state that the system spent the least amount of 
discrete time in.\footnote{In case of a tie-pick your favorite!} A data record
may contain one, two or more visits to the rare state. The data record may be
decomposed into cycles beginning and ending with visits to the rare state and
the following cycle statistics may be collected:

the average number of discrete visits to each state during a data record averaged cycle,
\begin{equation}\label{pi's}
\boldsymbol{\pi} = \lbrace \pi_{1}, \pi_{2},\ldots ,\pi_{N}\rbrace          
\end{equation}

the average rate of a visit for each state during a data record averaged cycle,
\begin{equation}\label{q's}
\boldsymbol{\mathsf{q}} = \lbrace \mathsf{q}_{1}, \mathsf{q}_{2},\ldots ,\mathsf{q}_{N}\rbrace.           
\end{equation} 
That is, $\mathsf{q}_{i}^{-1}$ is the continuous time length of a typical visit to the $i^{th}$
state.

The Carlson depth in this setting is the continuous time length of the characteristic cycle between
visits to the rare state given by
\begin{equation}\label{CD}
C\!D =  \frac{\pi_{1}}{\mathsf{q}_{1}} + \frac{\pi_{2}}{\mathsf{q}_{2}} +\ldots +  \frac{\pi_{N}}{\mathsf{q}_{N}}.          
\end{equation}

All of the quantities introduced so far are, in principle, easily computed from a sufficiently
detailed data record. On the other hand,  of equal interest to the theoretically minded observer 
is the the Hamiltonian of the 
subsystem and the temperature of the bath it is connected to(by way of fluctuation
exchanges). In what follows, it is supposed that energy and temperature have the same units, i.e.
that the Boltzmann $``k"$ has been absorbed into the temperature parameter. At this point the notions 
of energy and temperature are taken to be primitives but as the development proceeds
that will begin to change.

To each state $\upsilon(i) \in \boldsymbol{\upsilon}$ let there be an energy assignment $H_{i}$. Introduce
             
\begin{equation}\label{energy}
\mathbf{H} = \lbrace H_{1}, H_{2},\ldots ,H_{N}\rbrace           
\end{equation} 

The assignments in (\ref{energy}) are taken to be real numbers with units of
energy and serve as the Hamiltonian for the subsystem. The entire subsystem
is assigned a single temperature value, denoted by the symbol $\theta$.
$\theta$ takes its numerical values in the positive real numbers. It
carries units of energy.

 The following constraint ensures that the changes made to the Hamiltonian do not include 
 uniform additive shifts in reference point(which will expressed as temperature changes),
\begin{equation}\label{zerosum} 
\Sigma H_{k}=0.\footnote{The fixed choice of energy reference. This state of affairs is implicit 
in the Ising model, for a familiar example.}
\end{equation}

\subsection{Two Great Measures and Ergodicity}

While pursuing implications of the Birkhoff--Von Neumann quasi-ergodic
hypothesis is one of the chief goals of this paper a detailed
dissection of the argument is not. For the purposes of continuity
in the storyline
it is simply noted that according to the Liouville Theorem the fraction
of time spent  in the neighborhood of a phase point along the trajectory
is proportional to the volume of the neighborhood, which is preserved
by the motion.
That is, it is suggested that two measures are at play and that the
observations contained in a data record using either language should 
be directly comparable.

The first measure is based on the (relative)amount of time spent in particular
region of the phase space. The second is based on the (relative)volume of the region. 
As is well known, in the case of thermal equilibrium, the volume based measure
takes on the form of Gibbs-Boltzmann fame in which the dependency on the energy
of the region of phase space and temperature of the bath is made explicit.

In accord with these principles, in the context considered in this paper,
two measures are taken to be directly comparable. 

The first measure is the amount of time the observed system 
spent in a particular state $\upsilon(k) \in \boldsymbol{\upsilon}$
relative to the total length of the data record.
In the terminology of equation(\ref{CD})

\begin{equation}\label{timeprob} 
\textrm{Prob}(\upsilon(k))=\frac{\frac{\pi_{k}}{\mathsf{q}_{k}}}{\textrm{\footnotesize{CD}}}.
\end{equation}

The second measure is the microcanonical one, i.e.
the relative volume of the region in phase space,

\begin{equation}\label{energyprob}
\textrm{Prob}(\upsilon(k))= \frac{e^{- \frac{H_{\! k}}{\! \theta}}}{Z}
\end{equation}

where,

\begin{equation}
Z = e^{- \frac{H_{\! 1}}{\! \theta}}+e^{- \frac{H_{\! 2}}{\! \theta}}+\cdots+e^{- \frac{H_{\! N}}{\! \theta}}.
\end{equation}

A first easy consequence of (\ref{timeprob}), (\ref{energyprob}), and (\ref{zerosum}) is              

\begin{equation}\label{H.} 
H_{k}=\theta \, \log \Big( \,\frac{\Pi}{ x_{k}} \,\Big)
\end{equation}

where,
\begin{equation}\label{intro_x} 
x_{k} = \frac{\pi_{k}}{\mathsf{q}_{k}}
\; k=1,2,\ldots,N
\end{equation}

\begin{equation}
\Pi = \Big( x_{1}x_{2} \cdots x_{N} \Big) ^{\frac{1}{N}}
\end{equation}

and the $x_{k}$
take values in the positive real numbers and carry 
dimensions of time.

Equation (\ref{H.}) is an inspiration. It suggests that, within the range
of validity of the conventions and conditions adopted, more is contained
in the experimentalist's data record than mere statistics. The path itself
is described there and perhaps implicit in the time and frequency description
of the path lies the energy and temperature description. If only $\theta$
were known as a function of the ($x_{1}, x_{2}, \ldots, x_{N}$) equation
 (\ref{H.}) would deliver the state energies.
 
 It is perhaps somewhat surprising at the outset, that there is so very little ambiguity
 in what the function that maps $\theta$ to ($x_{1}, x_{2}, \ldots, x_{N}$)
 must be. In fact, figure 1 at the end of the next section, is a perspective of a constant temperature surface
 in time for a subsystem with three states, i.e. the case N = 3.\footnote{Constant temperature surfaces
 may be computed for arbitrary finite N but it is tough to produce the figure! Hence, N = 3.}

The next section develops machinery to seek out surfaces of constant temperature in time. 
Before leaving this section, please note
that a dilatation of the energy temperature space, i.e. the mapping

\begin{displaymath}\label{energydilatation}
    (\mathbf{H}; \theta) \rightarrow  (\mathbf{H} + \epsilon \mathbf{H}; \theta + \epsilon \theta)
\end{displaymath}
where $\epsilon$ is the real valued dilatation parameter, leaves the probabilities
    in (\ref{energyprob})  invariant.

Similarly, a dilatation of the time coordinates ($x_{1}, x_{2}, \ldots, x_{N}$)
from (\ref{intro_x}) leaves
the probabilities described in (\ref{timeprob}) invariant. That is, probabilites
are constant along rays in either space. A moments reflection reveals that
dilatation is the only mapping that leaves the pointwise probabilities invariant.
\footnote{ For example, a rotation may leave the total probability
of a set unchanged(for example a cone symmetric about the axis of rotation) but the 
N probabilities of any single point in the  set will not be invariant under the rotation.
}

\section{ Matched Invariants Principle}

A single experiment is performed and two observers are present and taking notes.
The first observer creates a data record from which the quantities referenced
in (\ref{pi's}) and (\ref{q's}) may be computed. The second observer creates
a data record in such a way that the quantities in (\ref{energy}) and $\theta$
are readily computed. The goal is to discover the method of observer number two.

The output of a single experiment then is two data points(one per observer): a single point in 
in the ($x_{1}, x_{2}, \ldots, x_{N}$) space and a single point in the $(H;\theta)$ space.

In the event that another  experiment is performed and the observers repeat the
activity of the previous paragraph, the  data points generated are either both the same
as the ones they produced as a result of the first experiment
or else both are different. This is the essence of the principal-the two observers
are using different languages to describe a $\it{single}$ event.
It is convenient to think that a series of experiments are under observation.
The sequence of data points generated by an observer traces out a curve.
There will be one curve in each space.  
It is assumed that the curves share a single, common parameterization.
That is, that the two observers share the same clock. 

$\it{The \; principle \; follows}$: in terms of probabilites, the two observers will continue to 
produce consistent results in the special case where the data points
in their respective spaces have changed from the first experiment to the second
but the probabilites have not. That is, if one observer experiences a dilatation
so does the other.

Of course, if the observers are able to agree if dilatation has occurred they are also able to agree
that it has not. In particular, in terms of differential displacements of the data points, the observers 
are able to decompose
a displacement into a components parallel to the dilatation direction
and perpendicular to it. Further, the observers are able to agree which component is which.

A good local set of coordinates for the observers consists of the dilatation
direction (that's one dimension so far...)  and any N-1 of the N probability
gradients, $\nabla P_{k}, \; k= 1,2 \ldots,N$. A concrete example of what
these gradients are and how they are applied to compute constant temperature
surfaces via  the matched invariants principle
will be developed in the next few subsections.

It is useful to note that, in either space, all probability gradients are 
perpendicular to the direction of dilatation. 
\footnote{Pointwise probabilities are invariant under dilatation}
As a consequence of the matched invariants principle, the two observers can agree 
on the event that the computed probabilities 
change and that
the asociated trajectories in thier respective spaces are perpendicular to the
dilatation direction at each point. 
\footnote{If a trajectory may be expressed as a sequence of displacements
each displacement perpendicular to the dilatation direction then the trajectory
lies on the surface of a sphere. That is, the differential relation
\begin{equation}
\mathbf{x} \cdot d \mathbf{x} = 0
\end{equation} defines the surface of the sphere.
}

 Summary: The probabilty invariant direction is the dilatation
direction. In the setting of a system with N possible states,
the N-1 dimensional space perp to the dilatation is spanned by 
any  set of N-1 probability gradients.

\subsection{ Comparison of Trajectories Perpendicular to the Dilatation Direction}

As was foretold, the ultimate purpose of this subsection 
is to apply the matched invariants
principle and develop the machinery that will allow observer number one to compute 
constant $\theta$ surfaces in ($x_{1}, x_{2},\ldots, x_{N}$) space.

For the purposes of keeping those calculations straightforward, albeit at the
cost of some clarity in the theoretical development, a specific reference frame
is introduced and eventually a specific number of dimensions will be chosen. 
It is hoped that the symmetry of the result, displayed in the
figure 1, will serve a posteriori to boost the level of theoretical
clarity.

In what follows observer number one chooses coordinates
($x_{1}, x_{2},\ldots, x_{N}$)
\footnote{Recall equation (\ref{intro_x})} 
and observer two
chooses independent coordinates
($H_{1}, H_{2},\ldots, H_{N - 1}, \theta$).
\footnote{Recall equation (\ref{zerosum}).}

Local trajectories are decomposed into components along the probability
gradients and along the dilatation direction. The form of these gradients
is displayed next.

On the observer one side, consider $p_{1}$  
for example:
\footnote{Recall equations (\ref{timeprob}) and
(\ref{intro_x}).
Derivatives have been taken with respect to
the coordinates $x_{1}, x_{2},\ldots, x_{N}$. }

\begin{equation}\label{gradp1dat}
 \nabla p_{1} = \frac{1}{C \!D}\left\{ \begin{array}{ll}
                        1& \\ 0 &  \\ \vdots & \\ 0
            \end{array} 
 \!\! \!\! \!\!   \right\} -
    \frac{p_{1}}{C \!D}\left\{ \begin{array}{ll}
                        1& \\ 1 &  \\ \vdots & \\ 1
            \end{array} 
  \!\! \!\! \!\!  \right\}
\end{equation}

Note that $\nabla p_{1} \cdot \mathbf{x} =0$.

Meanwhile, on the observer two side:
\footnote{Recall (\ref{energyprob}). Derivatives
have been taken with respect to $H_{1}, H_{2},\ldots, H_{N - 1}, \theta$. }

\begin{equation}\label{gradp1eng}
 \nabla \mathsf{p}_{1} = \mathsf{p}_{1} \left\{ \begin{array}{ll}
                      \, - \frac{1}{\theta}&  \\ \vphantom{o} & \\ \quad 0 &  \\ \quad \vdots & \\ \quad 0 &
                      \\ \vphantom{o} &  \\ \,\, \frac{\sf{H}_{1}}{{\theta}^2}
            \end{array} 
    \right\} -
                \mathsf{p}_{1}\left\{ \begin{array}{ll}
                        \frac{\mathsf{p}_{N}-\mathsf{p}_{1}}{\theta}& \\  \vphantom{o} & \\ \frac{\mathsf{p}_{N}-\mathsf{p}_{2}}{\theta} &  \\ \ \quad \vdots & \\ 
                        \frac{\mathsf{p}_{N}-\mathsf{p}_{N-1}}{\theta} & \\ \vphantom{o} & \\ \quad \frac{\sf{U}}{{\theta}^2}
            \end{array} 
  \! \! \! \! \! \!\! \!\! \right\}
\end{equation}

Note that $\nabla \mathsf{p}_{1} \cdot (\sf{H};\theta) =0$.

Beyond dilatation, another thing that observers can agree upon is 
the magnitude of changes in the state  probabilities.
\footnote{See the discussion surrounding ergodicity and the direct comparison
of (\ref{timeprob}) and (\ref{energyprob}).} 
In particular, probability changes that arise due to transitions in the plane perpendicular 
to the
dilatation direction can be agreed upon.
Recall that this space is spanned by any N-1 of the probability gradients.

Consider a specific example, the case where N = 3. A 
finite size step, $\Delta \mathbf{x}$,
(suitable for use in a numerical scheme) 
perpendicular to the dilatation 
direction has the form

\begin{equation}\label{da_for_data} 
\Delta \mathbf{x} = \Delta_{1} \; \frac{\nabla p_{1}}{ \sqrt{ \nabla p_{1} \cdot \nabla p_{1} } }
+ \Delta_{2} \; \frac{\nabla p_{2}}{ \sqrt{ \nabla p_{2} \cdot \nabla p_{2} } }
\end{equation} 

where $\Delta_{1}$ and $\Delta_{2}$ are small numbers.
That is, (\ref{da_for_data}) is an experimentalist's approximation
of the mathematician's differential. Once limits are taken, the change in
probabilities incurred by the displacement (\ref{da_for_data}) are given by:

\begin{equation}\label{dp1data} 
d p_{1} = \nabla p_{1} \cdot d\mathbf{x}
\end{equation} 
\begin{equation}\label{dp2data} 
d p_{2} = \nabla p_{2} \cdot d\mathbf{x}
\end{equation} 

Similar equations hold in the energy space.
In the case where $N = 3$, the general displacement perpendicular
to the direction of dilatation is lies
in the space spanned by any two of the probability gradients in that space.
In  ($H_{1}, H_{2},\ldots, H_{N - 1}, \theta$) coordinates, 
observer number two's analog of equation (\ref{da_for_data}) is a linear combination of those
gradients

\begin{equation}\label{da_for_energy} 
(\delta\mathbf{\sf{H}}, \delta\theta) = \delta_{1} \; \frac{\nabla \mathsf{p}_{1}}{ \sqrt{ \nabla \mathsf{p}_{1} \cdot \nabla \mathsf{p}_{1} } }
+ \delta_{2} \; \frac{\nabla \mathsf{p}_{2}}{ \sqrt{ \nabla \mathsf{p}_{2} \cdot \nabla \mathsf{p}_{2} } }.
\end{equation} 

The second observer's analog of (\ref{dp1data}) and (\ref{dp2data}) are
\begin{equation}\label{dp1eng} 
d \mathsf{p}_{1} = \nabla \mathsf{p}_{1} \cdot (d\mathbf{\sf{H}}, d \theta)
\end{equation} 
\begin{equation}\label{dp2eng} 
d \mathsf{p}_{2} = \nabla \mathsf{p}_{2} \cdot (d\mathbf{\sf{H}}, d \theta)
\end{equation}

Observers
agree on the probability changes seen so that:

\begin{equation}\label{dp1s_equal} 
d \mathsf{p}_{1} = d p_{1}
\end{equation}
\begin{equation}\label{dp2s_equal} 
d \mathsf{p}_{2} = d p_{2}
\end{equation}

Importantly, equations (\ref{dp1s_equal}) and (\ref{dp2s_equal}) provide a relation between
the $\Delta_{.}$'s from equation  (\ref{da_for_data})  and the $\delta_{.}$'s
from equation (\ref{da_for_energy})

\begin{equation}\label{deltas_from_dPs}
 \left(\begin{array}{cc}
\delta_{1} & \\ 
\vphantom{X}\\
\delta_{2} 
\end{array} \!\!\!\!\!\!\right) = 
\left(\begin{array}{cc}
\Vert \nabla \mathsf{p}_{1} \Vert & \frac{\nabla \mathsf{p}_{1} \cdot \nabla \mathsf{p}_{2}}{\Vert \nabla \mathsf{p}_{2} \Vert}\\
\vphantom{X}\\
\frac{\nabla \mathsf{p}_{2} \cdot \nabla \mathsf{p}_{1}}{\Vert \nabla \mathsf{p}_{1} \Vert} & \Vert \nabla \mathsf{p}_{2} \Vert
\end{array}\right)^{-1} 
\left(\begin{array}{cc}
d \mathsf{p}_{1} & \\
\vphantom{X}\\ 
d \mathsf{p}_{2} 
\end{array} \!\!\!\!\!\!\ \right).      
\end{equation} 

Equation (\ref{deltas_from_dPs}) is the relation that observer one was looking for.
If observer two supplies the $\delta_{.}$'s observer one can translate them into
$\Delta_{.}$'s using (\ref{deltas_from_dPs}), (\ref{dp1s_equal}), (\ref{dp2s_equal}),
(\ref{dp1data}), (\ref{dp2data})
and (\ref{da_for_data}). That is, changes in $\theta$ can be translated into changes
in the time coordinates.

A consequence of (\ref{dp1s_equal}) and (\ref{dp2s_equal}) under the agreement that the 
observers share the same clock is

\begin{equation}\label{dotp1s_equal} 
\dot{\mathsf{p}_{1}} = \dot{ p_{1}}
\end{equation}
\begin{equation}\label{dotp2s_equal} 
\dot{\mathsf{p}_{2}} = \dot{p_{2}}
\end{equation}

where the dot denotes differentiation with respect
to the single, common curve parameter.

\subsection{ Consequence of Matched Invariants Principle}
Another consequence of the mathched invariants principle is that, as one
moves perpendicularly to the dilatation direction from one point to the next along the surface
of a sphere in the ($x_{1}, x_{2},\ldots, x_{N}$) space,
the ratio of temperature changes along the ray (the dilatation of temperature) remains 
constant.\footnote{The discussion which follows
 keeps the matched invariants principle to the fore.
A simpler geometrical demonstration (obscuring the notion of observation of probabilities) 
goes as follows. Let two points on the 
$\theta$-axis be labelled $\theta$ and $\theta'$. Consider a rigid rotation 
in the ($H_{1}, H_{2},\ldots, H_{N - 1}, \theta$) space (L2 moduli
of the original line segments on the theta axis maintained). The temperature at the rotated images are easily
obtained by vector dot product with the theta axis. In the ratio the cosines disappear and the 
original ratio of the L2 moduli is maintained.
}

Further aspects of the example, N = 3, from the 
previous section are developed in this subsection.
Consider two nearby points along the same ray, $r_{initial}$ in ($x_{1}, x_{2}, x_{3}$) space:
$ A =\{ x_{1}, x_{2},  x_{3}\}$ and $A' =\{ x_{1} + \alpha x_{1}, x_{2} + \alpha x_{2},  
x_{3} + \alpha x_{3}\}$
where $\alpha << \textrm{\small{CD}}$. Since the two points are on the same ray they generate the same probabilities
(as do their images in $(\mathbf{H};\theta)$ space) and in particular

\begin{equation}\label{rayconstant1} 
\frac{ \sf{H}_{\sf{k}} - \sf{U} }{\theta}(A) =  \frac{ \sf{H'}_{\sf{k}} - \sf{U'} }{\theta'}(A')
\end{equation}

Suppose points A and $A'$ on ray $r_{initial}$ undergo a small displacement 
$A \rightarrow B$ and $A' \rightarrow B'$ perpendicular to $\textrm{r}_{initial}$
(the direction of dilatation of the two points). Further suppose that points were displaced in such a way
that the displaced points $B$  and $B'$  both lie on a ray, $\textrm{r}_{final}$.

Denote the temperature at A by $\theta$ and at $A'$ by $\theta'$.
The temperature at B is given according to equations (\ref{da_for_energy}) and
(\ref{gradp1eng}) by

\begin{equation}\label{theta_at_B} 
\theta + \Big[ \mathsf{p}_{1}  \frac{ \sf{H}_{1} - \sf{U}}{ \theta } \Big] \; \frac{1}{\theta} \frac{\delta_{1}}
{\Vert \nabla \mathsf{p}_{1} \Vert}
+ \Big[ \mathsf{p}_{2}  \frac{ \sf{H}_{2} - \sf{U}}{ \theta } \Big] \;  \frac{1}{\theta} \frac{\delta_{2}}
{\Vert \nabla \mathsf{p}_{2} \Vert}
\end{equation} 

The quantities in square brackets above are invariant along the ray and thus are the same at B and at $B'$.
The scaling of the probability gradients along the ray from point A to $A'$ may be observed from (\ref{gradp1eng}), 
for example as follows

\begin{equation}\label{gradP_scales_along_ray} 
\nabla \mathsf{p}_{1} \Big\vert_{A} = \; \frac{\theta '}{\theta} \, \nabla \mathsf{p}_{1} \Big\vert_{A'} 
\end{equation}

The scaling of the $\delta_{K}$ is obtained from equation (\ref{deltas_from_dPs})
and the requirement 
that both B and $B'$ lie on $\textrm{r}_{final}$, 
i.e. that the probability changes
incurred along path $A \rightarrow B$ are the same as those along $A' \rightarrow B'$.
For example, in the case of the first coordinate,

\begin{equation}\label{delta_scales_along_ray} 
\delta_{1} \Big\vert_{A \rightarrow B} = \; \frac{\theta}{\theta'} \,\,  \delta_{1}' \Big\vert_{A' \rightarrow B'} 
\end{equation}

Thus the temperature at $B'$ is obtained from equations (\ref{theta_at_B}), (\ref{gradP_scales_along_ray}) and 
(\ref{delta_scales_along_ray}) is found to be

\begin{equation}\label{theta_at_B'} 
\frac{\theta'}{\theta}
\Big( \theta + \Big[ \mathsf{p}_{1}  \frac{ \sf{H}_{1} - \sf{U}}{ \theta } \Big] \; \frac{1}{\theta} \frac{\delta_{1}}
{\Vert \nabla \mathsf{p}_{1} \Vert}
+ \Big[ \mathsf{p}_{2}  \frac{ \sf{H}_{2} - \sf{U}}{ \theta } \Big] \;  \frac{1}{\theta} \frac{\delta_{2}}
{\Vert \nabla \mathsf{p}_{2} \Vert} \Big)
\end{equation} 

Comparison of (\ref{theta_at_B}) and (\ref{theta_at_B'}) reveals that the original ratio
of temperatures at points A and $A'$ on ray $\textrm{r}_{initial}$ is maintained by the motion
$A \rightarrow B$ and $A' \rightarrow B'$ described above.

\subsection{ The Sense of Direction and Choice of Scale of the Theta to Centerline Map}

Observer Two reports that along the $\theta$ axis in ($H_{1}, H_{2},\ldots, H_{N - 1}, \theta$) space
the probabilities in (\ref{energyprob}) are uniform. Observer one reports
that the same is true of
the $\it{centerline}$ axis, i.e. ray from the origin
containing the  $\{1,1,\ldots, 1\}$ vector, in
($x_{1}, x_{2},\ldots, x_{N}$) space. 
This simple observation identifies the image of the theta axis
in the ($x_{1}, x_{2},\ldots, x_{N}$) space as the centerline.

Taking into account the results of the last two subsections, it is clear that
knowing the behaviour of $\theta$ along the centerline allows
Observer one to determine
the $\theta$ field in ($x_{1}, x_{2},\ldots, x_{N}$) space completely.

Recall, that an algorithm has been prescribed which gives the evolution of $\theta$
along a spherical surface in the ($x_{1}, x_{2},\ldots, x_{N}$) space given the value of $\theta$ at the 
intersection of the surface
with the centerline.

Once the $\theta$ values on a single spherical surface are obtained
the rest of the field is determined via the preservation of $\theta$ dilatation 
property discussed in the previous 
subsection.  Take the mapping of the $\theta$ axis
to the  centerline as the reference.

So what can be said about the centerline map?
Important features of any continuous, one to one, and onto map between one-dimensional
spaces include the sense of 
orientation (as one moves toward infinity along the centerline in ($x_{1}, x_{2},\ldots, x_{N}$)
space is $\theta$ going to infinity or is $\theta$ moving to zero?)
and for a given orientation, a choice of scale, i.e. the local stretching or shrinking of the
map.
 
For example,  the identity map $\theta(\textrm{\small{CD}}) = \textrm{\small{CD}}$ is a
choice of temperature scale
oriented such that theta goes to infinity as one moves out to infinity along the centerline. A simple inversion 
$\theta(\textrm{\small{CD}}) = \frac{1}{CD}$ is a choice of scale that takes theta to zero as one moves
out to infinity along the centerline.

Note that the sense of direction of the map may be obtained form the following thought experiment:
An experiment is conducted in a laborory frame that produces a set of set changes in time-a data flow.
One might think of this as a sort of information ``velocity'', the rate at which an observer becomes aware
that a state change has taken place.
An observer in the laboratory frame measures the Carlson depth of the data flow.\footnote{Recall (\ref{CD}).}

Now a second observer, moving at a constant ``velocity'' (with respect to the
laboratory frame) forward into the data flow, sees exactly
the same probabilities of events as the laboratory frame observer but witnesses the time inbetween
visits to the rare state to pass by more quickly than the observer in the lab frame.
In fact, the observer may move forward at arbitrarily large constant ``velocity'' and
witness, the same sequence of ordered events occurring with the same probability
as the laboratory frame observer but the Carlson depth shrinks to zero as the observer's
speed relative to the lab frame tends to infinity.

A third observer, moving more slowly than the observer in the laboratory frame, sees
the data pass by more slowly than the lab observer. But, importantly, this decrease in ``velocity''
relative to the lab observer may not increase without bound. A point is reach when the travelling 
observer is moving away from the data at the same rate at which the data is moving toward
her at which point: nothing changes. Identify this notion of total absence of change of
the state of the system with the notion of absolute zero.

Hence we adopt the convention, consistent with the principle of casuality, that as
the temperature is decreased to zero the recurrence time of the system is observed to 
increase and conversely. 

This convention prescribes the sense of orientation of the map. It does not suggest
a choice of scale. In fact, one conjectures that given many reasonable choices of scale
there is some one-dimensional parameterized family
of accelerating frames of observation that would produce results consistent with it.
Analysis of those structures is not considered
here.\footnote{The author is indebted to S. Huntsman for pointing out the
the analogy between the proceeding arguments and the Unruh-Davies Effect. }

\begin{figure}
\begin{center}
\epsfig{file=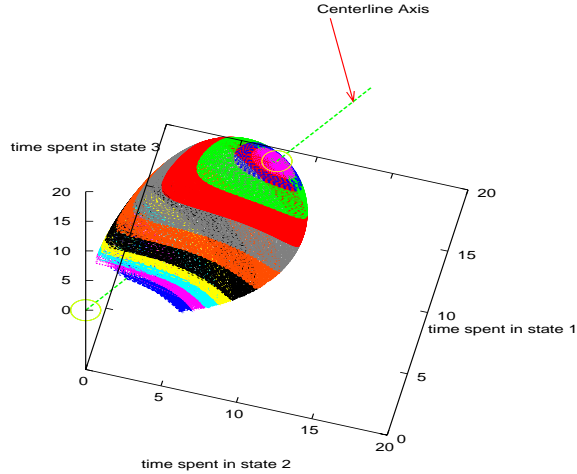,width =8cm, height =8cm}
\caption{Isothermal Surface Drawn in Space of Statewise
Contributions to Fundamental System Period for a Three State
System}
\end{center}
\end{figure}
%\begin{figure}
%\begin{center}
%\epsfig{file=fig3.eps, width =5cm, height =5cm}
%\caption{test test test}
%\end{center}
%\end{figure}

In the figure,
a $\theta$ to centerline mapping  $\theta(\textrm{\small{CD}}) = \frac{A}{CD}$,
where $A$ is a positive real constant,
was 
taken as the choice of scale along the centerline. Note that 
the centerline axis is perpendicular to the surface at their intersection.
Local to the centerline axis, the system period fixes the temperature
(within a choice of scale).
The reader may have their own favorite choice of temperture scale
to which the Matched Invariants Principle
may then be applied to generate the temperature field from which  
surfaces of constant $\theta$ (consistent with that choice of
scale) may be isolated.

\section{ Application of the Foregoing to Closed Queueing System}

In the previous sections, particular characteristics of the experiment generating the observed dynamics 
were never explicitly
stated. The subject never came up.
As a simple example of an application of the foregoing to a
non-standard system, in this section, some thermodynamical aspects 
associated with the dynamics of a
closed queueing network are derived.

Necessary fundamentals from elementary queueing theory and elementary thermodynamics
are introduced, as required, along the way. The following queue construction is
standard.\footnote{See for example, Kleinrock, L.  $\it{Queueing \; Systems}$,
Wiley, New York, 1975.} Connections with the "Stosszahlansatz" may be
found in work of Kac, Uhlenbeck and Siegert.\footnote{See for example, A.J Siegert, $\it{On \; the \; Approach \; to \;
Statistical \; Equilibrium}$,The Physical Review, v76, 1949.}

Recall that a closed queueing system consists of N buckets and M balls. Traditionally the
N buckets represent  N servers say at the office of some goverment bureaucracy and the M balls
are clients waiting to be served. That is, they wait in line to get to the counter at which time
they are told they are in the wrong line and need to go to a different counter, hence closed.

At any moment in time an observer might see $m_{1}$ people at counter $1$ 
(that is, $m_{1}$ balls in bucket $1$) and 
$m_{2}$ balls in bucket $2$
etc. The list of bucket occupancies
observed, denoted by $(m_{1},m_{2},...,m_{N})$, determines the instataneous state of the system. 
An instantaneous state of the system will herein be referred to as a 
$\it{configuration}$. Let
$\boldsymbol{\upsilon}$ denote the set, indexed by
the natural numbers, of all possible system configurations.

When the queue is observed in action, the list of configurations it visits in 
time (a list of lists indexed by time) forms the experimental record. 
It is the 
trajectory of the system as it moves through its state space.

Once a trajectory is in hand the statistics in (\ref{pi's}), (\ref{q's}) 
and (\ref{intro_x})
may be collected. 
Further, it is known from the previous sections that there is only one way, consistent
with the Matched Invariants Principle and the observers choice of temperature scale,
to define a temperature field in ($x_{1}, x_{2}, \ldots, x_{N}$) space where the 
computed statistics of the trajectory reside. By way of  (\ref{H.}), the vector field of
energies is generated next.

In the context of a specific system, like the present example, one can go further.
It makes sense to look for work and heat. 

Recall from elementary thermodynamics that the macroscopic observable energy
is defined by

\begin{equation}\label{Workstart} 
U =\mathsf{p} \cdot \mathbf{H}
\end{equation} 

where $\mathsf{p}$ is given by (\ref{energyprob})
and $\mathbf{H}$ by (\ref{energy}).

Take a derivative with repect to time to obtain

\begin{equation}\label{energyrate} 
\dot{U} = \dot{\mathsf{p}}\cdot \mathbf{H} + \mathsf{p}\cdot \dot{\mathbf{H}}.
\end{equation}

In the equilibrium setting, the RHS of  (\ref{energyrate}) decomposes the rate of
energy change into work rate and heat rate respectively.

That is to say the work rate is given by

\begin{equation}\label{Workrate} 
\dot{W} = \mathsf{p}\cdot \dot{\mathbf{H}}
\end{equation}

and the heating rate by

\begin{equation}\label{heatrate} 
\dot{Q} = \dot{\mathsf{p}} \cdot \mathbf{H}
\end{equation}

\subsection{Compute  $ \mathsf{p} \dot{\mathbf{H}}$}

Now the two vectors appearing in the dot product in the RHS of (\ref{Workrate})
are elements of a high-dimensional space (the cardinality of $\upsilon$)while the work
rate itself is a single scalar quantity. So in switching form the RHS to the 
LHS in  (\ref{Workrate}) one also switches levels of description and goes from
too much information to perhaps not enough. Thermodynamic tradition insists
on the introduction of an alternative, a middle
ground between information overload and ignorance.

What is required are two low dimensional vectors (low in comparison to the
the number of microscopic states $\boldsymbol{\upsilon}$ represents) that contain sufficient information
to characterize the observed trajectory and are equivalent to $\mathsf{p}$ and $\dot{\mathbf{H}}$
in the sense that the inner product is the same in both sets of descriptions
of the dynamics. Famous examples of such conjugate pairs are the pressure and rate of volume change
in a gas and the average magnetization and applied field change in the case of a lattice
magnet. 

A good starting point is simply to write out exactly what the high-dimensional
form of the work rate looks like.

Recall from (\ref{H.})

\begin{equation}
H_{.}=\theta \, \log \Big( \,\frac{\Pi}{ x_{.}} \,\Big)
\end{equation}

where, 
             
\begin{equation}
\Pi = \Big( x_{1}x_{2} \cdots x_{N} \Big) ^{\frac{1}{N}}.
\end{equation}

Equivalently,

\begin{equation}
H_{.}=\theta \, \log \Big( \,\frac{\gimel}{ p_{.}} \,\Big)
\end{equation}

where, 
             
\begin{equation}\label{bigpi} 
\gimel = \Big( p_{1}p_{2} \cdots p_{N} \Big) ^{\frac{1}{N}}.
\end{equation}

Therefore

\begin{equation}
\dot{H_{.}}=\dot\theta \, \log \Big( \,\frac{\gimel}{ p_{.}} \,\Big)
+ \theta \Big( \frac{\dot\gimel}{ \gimel}  - \frac{\dot p_{.}}{ p_{.}}    \Big)
\end{equation}

and so

\begin{equation}\label{Workrateq} 
 \mathsf{p}\cdot \dot{\mathbf{H}} = \dot\theta \, \log \Big( \,\frac{\gimel}{ \Lambda} \,\Big)
 + \theta \Big( \frac{\dot\gimel}{ \gimel}     \Big)
\end{equation}

with             
\begin{equation}\label{biglambda} 
\Lambda = \Big( p_{1}^{p_{1}}p_{2}^{p_{2}} \cdots p_{N}^{p_{N}} \Big). 
\end{equation}

\subsection{In Particular for the Closed Queue}

The goal is to obtain a form of the work rate suitable for the 
closed queueing system. The calculations involved are simple
but some further details about the construction
of the queue are necessary before system specific
calculations can begin. 

A closed queue of M servers is constucted as a Markov process by characterizing
the holding times of the servers, the interaction among the servers and the state space of the queue.

The characteristic times of the M servers are the rates in the exponential distribution
that characterizes a particular servers processing time. Routing from one server to another
is governed by an $M \times M $ matrix
of transition probabilities,
herein called the routing table. A basic transition occurs as follows:
 at the completion of a service time the server looks to the routing table
 for instruction. The length of the service is at the discretion
 of the server. The location of the client's next service is
 decided by the router. 

It is well known that such a transition matrix admits an eigenvector with all positive
coefficients. Without loss of generality take the minimum component of the eigenvector
to have value 1(the rare state gets visited once).

If m is the total number of clients in the system
then the states of the queueing network are all M-long sequences of non-negative
integers(zero is ok) that sum to m. As is well known, that number is given by

\begin{equation}\label{numberofstates}
N= \binom{M + m - 1}{m}
\end{equation}

In summary then, the equilibrium distribution of the closed queueing network is given by specifying
the parameters for the exponentially distributed service times

\begin{equation}\label{que-q's} 
\mathbf{\sf{q}} = \lbrace (q_{1}, q_{2}, \cdots, q_{M}):  q_{.}
\in  \mathbf{R}^{+}  \rbrace. 
\end{equation}

The eigenvector, with eigenvalue $1$, of the $M \times M $ matrix
of probabilities (the routing table)

\begin{equation}\label{que-pi's} 
\mathbf{\pi} = \lbrace (\pi_{1}, \pi_{2}, \cdots, \pi_{M}):  \pi_{.}
\in  [ 1, \infty ) \; \textrm{and} \; \pi_{rare} = 1  \rbrace 
\end{equation}

and the state space

\begin{equation}\label{questates} 
\boldsymbol{\upsilon} = \lbrace (\upsilon_{1}, \upsilon_{2}, \cdots, \upsilon_{M}):  \upsilon_{.}
\in  \mathbf{Z}^{+} ; \upsilon_{1}+ \upsilon_{2} + \cdot + \upsilon_{M} = m \rbrace. 
\end{equation}

The equilibrium distribution for a closed queue is given by

\begin{equation}\label{queProb} 
\mathbf{Prob}(\alpha_{1}, \alpha_{2}, \cdots, \alpha_{M}) = 
\frac{
( \frac{\pi_{1}}{q_{1}})^{\alpha_{1}} (\frac{\pi_{2}}{q_{2}})^{\alpha_{2}} 
\cdots (\frac{\pi_{M}}{q_{M}})^{\alpha_{M}}
} 
{  
\sum_{\upsilon} ( \frac{\pi_{1}}{q_{1}})^{\upsilon_{1}} (\frac{\pi_{2}}{q_{2}})^{\upsilon_{2}} 
\cdots (\frac{\pi_{M}}{q_{M}})^{\upsilon_{M}}
}.  
\end{equation}

In this setting, the quantity $\gimel$ from (\ref{bigpi}) takes a particularly pleasing form

\begin{equation}\label{queProbgimel} 
\gimel=
 \frac{ \Big[
 \frac{\pi_{1}}{q_{1}} \frac{\pi_{2}}{q_{2}} 
\cdots \frac{\pi_{M}}{q_{M}} \Big]^{\frac{\alpha}{N}}
} 
{  
\sum_{\upsilon} ( \frac{\pi_{1}}{q_{1}})^{\upsilon_{1}} (\frac{\pi_{2}}{q_{2}})^{\upsilon_{2}} 
\cdots (\frac{\pi_{M}}{q_{M}})^{\upsilon_{M}}
}  
\end{equation}

where
\begin{equation}\label{alpha}
\alpha = {\sum_{k=0}}^{m-1} (m-k) 
\binom{(M - 1) + k - 1}{(M-1) - 1}. 
\end{equation}

Note that
\begin{equation}
 \frac{\alpha}{N} = \frac{m}{M}.
\end{equation}

The heart of the right hand side of (\ref{Workrate}) for the queue involves the particular expression

\begin{equation}\label{almostwork} 
 \frac{\dot\gimel}{ \gimel} = \frac{\overset{\centerdot}{\Big[
 \frac{\pi_{1}}{q_{1}} \frac{\pi_{2}}{q_{2}} 
\cdots \frac{\pi_{M}}{q_{M}} \Big]^{\frac{\alpha}{N}}}}
{\Big[
 \frac{\pi_{1}}{q_{1}} \frac{\pi_{2}}{q_{2}} 
\cdots \frac{\pi_{M}}{q_{M}} \Big]^{\frac{\alpha}{N}}} \; - \;
\overset{\centerdot}{
\frac{
      \Big[
      \sum_{\boldsymbol{\upsilon}} 
      ( \frac{\pi_{1}}{q_{1}})^{\upsilon_{1}} 
      (\frac{\pi_{2}}{q_{2}})^{\upsilon_{2}} 
      \cdots 
      (\frac{\pi_{M}}{q_{M}})^{\upsilon_{M}} 
      \Big]
      }
     { 
       \Big[
       \sum_{\boldsymbol{\upsilon}} 
       ( \frac{\pi_{1}}{q_{1}})^{\upsilon_{1}} 
       (\frac{\pi_{2}}{q_{2}})^{\upsilon_{2}} 
       \cdots 
       (\frac{\pi_{M}}{q_{M}})^{\upsilon_{M}}
       \Big]
}
}.
\end{equation}

After straightforward calculation this simplifies to

\begin{multline}\label{workcore}
 \frac{\dot\gimel}{ \gimel} = \frac{\alpha}{N} \bigg(
 \frac{\dot{ [ \frac{\pi_{1}}{q_{1}} }]}{\frac{\pi_{1}}{q_{1}}}
   +
   \frac{\dot{[ \frac{\pi_{2}}{q_{2}} }]}{\frac{\pi_{2}}{q_{2}}}
   +
\cdots \frac{\dot{ [\frac{\pi_{M}}{q_{M}} }]}{\frac{\pi_{M}}{q_{M}}}\bigg)\\
- 
\bigg(
 \frac{\langle \upsilon_{1} \rangle 
 \dot{ [ \frac{\pi_{1}}{q_{1}} }]}{\frac{\pi_{1}}{q_{1}}}
   +
   \frac{\langle \upsilon_{2} \rangle\dot{[ \frac{\pi_{2}}{q_{2}} }]}{\frac{\pi_{2}}{q_{2}}}
   +
\cdots \frac{\langle \upsilon_{M} \rangle\dot{ [\frac{\pi_{M}}{q_{M}} }]}{\frac{\pi_{M}}{q_{M}}}\bigg),
\end{multline}
where the angle brackets denote the average is taken over the set (\ref{questates}) with respect to
measure (\ref{queProb}).

First of all, note that the description of the dynamics appearing
on the RHS of (\ref{workcore})is low dimensional. Recall that M is the number of buckets
and compare this with (\ref{numberofstates}). 
Equally important, the pieces comprising the right hand side of (\ref{workcore}) are easily recognizeable.

Let $\Upsilon$ denote the average state of the system over an observation period,

\begin{equation}\label{bigupsilon}
\Upsilon = \mathsf{p} \cdot \boldsymbol{\upsilon}
\end{equation}

that is, the average ball count per bucket. 

Let $\boldsymbol{\eta}$ contain the performance information for each the M servers
in the form of M ratios:  average number of customers 
assigned to that server per characteristic
cycle scaled by the servers characteristic rate

\begin{equation}
\boldsymbol{\eta} = \Big\{ \frac{\pi_{1}}{q_{1}} ,\frac{\pi_{2}}{q_{2}} ,\ldots,\frac{\pi_{M}}{q_{M}} \Big\}.
\end{equation} 

In terms of these quantities the workrate
takes the form

\begin{equation}\label{Workfinal} 
\dot{W} = \frac{\dot{\theta}}{\theta} \; U + \left( \frac{\alpha}{N} \mathbf{1} - \Upsilon \right) \cdot 
\frac{\dot{\boldsymbol{\eta}}}{\boldsymbol{\eta}}. 
\end{equation} 

In the event that the work takes place at constant temperature we have simply

\begin{equation}\label{Workfinalnotheta} 
\dot{W} =  \left( \frac{\alpha}{N} \mathbf{1} - \Upsilon \right) \cdot 
\frac{\dot{\boldsymbol{\eta}}}{\boldsymbol{\eta}} 
\end{equation}

The macroscopic work rate in the RHS  of (\ref{Workfinalnotheta} )does all the things we need it to. 
The number of dimensions of the vectors involved have been greatly
reduced and the M-dimensional vectors obtained represent meaningful coarse
characteristics of the system. The dot product, i.e the scalar value
of the work rate, has been preserved. The first law of 
thermodynamics is observed by definition. What about the second law?

The second law question will be the subject of the next subsection.
Before getting there two additional famous thermodynamic quantities are
presented for this system. The calculations are similar to
the example just given and will be omitted.

The Helmholtz Free Energy:

\begin{equation}
F = - \log \Big( \sum_{\gamma \in \boldsymbol{\upsilon}}( \frac{\pi_{1}}{q_{1}})^{\gamma_{1}- \frac{\alpha}{N}}
( \frac{\pi_{2}}{q_{2}})^{\gamma_{2}- \frac{\alpha}{N}} \cdots
( \frac{\pi_{M}}{q_{M}})^{\gamma_{M}- \frac{\alpha}{N}}\Big)
\end{equation}

The Internal Energy:

\begin{equation}
U = \theta \Big\langle \; \log \Big( ( \frac{\pi_{1}}{q_{1}})^{\upsilon_{1}- \frac{\alpha}{N}}
( \frac{\pi_{2}}{q_{2}})^{\upsilon_{2}- \frac{\alpha}{N}} \cdots
( \frac{\pi_{M}}{q_{M}})^{\upsilon_{M}- \frac{\alpha}{N}}\Big) \; \Big\rangle
\end{equation}
where the angle brackets denote the average is taken over the set (\ref{questates}) with respect to
measure (\ref{queProb}).

\subsection{$2^{nd}$ Law for the Closed Queue}

The purpose of last subsection was to gather together 
and present a few famous thermodynamical highlights of this 
queue system.

Among these highlights however there is one quantity that is
more than just a mere calculation.
This quantity is the the least upper bound for the heating rate
of the queue
and is a specific example of what is
considered, in its abstract form, to be at the very heart of the Second Law. 

If a least
upper bound for the system can be identified concretely
then it does not need a Law to justify its exisitence.
It is a simply a matter of proving that an inequality is sharp.

In this particular system, the existence of a least upper bound for the
heating rate is very nearly obvious. The proof is simply
a matter of making the right definitions. Perhaps the following
heuristic argument for sharpness will suffice.
Consider that there are two probabilities at play: (\ref{timeprob})
and (\ref{queProb}).

In equilibrium\footnote{Equilibrium
in this context is taken to mean that the routing table
values and service rates are constant in time,}, the two measures 
are the same, equal. But what happens when the queue 
transitions to a new equilibrum? Suppose, for example,  that the 
routing table is changed very slowly in such a way that no work is 
performed\footnote{The work rate is identically zero throughout
the transition.}
and that the energy is increased.

Since the routing table has changed, its eigenvector, (\ref{que-pi's}) may also change
and so also the probabilities given by description (\ref{queProb}). The
probabilities given in (\ref{timeprob}) will eventually converge
to the the new (\ref{queProb}). 

An observer
watching the routing table needs to see each altered element of the $M \times M$ 
table route quite a few customers before being satisified that the new table values
have been empirically discovered and the new equilibrium achieved.

However, an observer watching the state space $\boldsymbol{\upsilon}$ needs to
see each of the N states, recall (\ref{numberofstates}), transition
many times before being satisifed that the probabilities of each
microscopic state given in  (\ref{timeprob}) have been empirically discovered.

Note that a microstate transition cannot occur without
a routing table action, i.e. the two ``step'' in time
at exactly the same scale but that, in practice, the number 
$N$ dwarfs the number $M$ squared.
 
In this situation, it is not
unreasonable to suppose that the observer of the routing table will 
become empirically satisifed that the table entries 
have converged to their new values and that the new energy
level has been attained long before the 
$\boldsymbol{\upsilon}$ observer sees each microstate in the 
state space visited even once,
let alone visited sufficiently many times that there is some empirical sense of convergence.
\footnote{In short, the ``covering time of the routing table'' is much, much smaller than the covering
time of the state space: the steps of the covering walks occur at identical rates (in fact they step at
exactly the same time, every time), the graphs are comparable(the routing table
graph is approximately a quotient space of the set of edges of the 
$\boldsymbol{\upsilon}$ graph, mod out by ``direction'')  but obviously one,
the $\boldsymbol{\upsilon}$ graph, is
much much bigger.}

Still one might imagine, along with Carnot, a series of very small perturbations of the routing table
with very long periods of time inbetween such that the two descriptions  (\ref{timeprob})
and (\ref{queProb}) are very nearly equal at all times. 

The heating upper bound proposed for this system is given by

\begin{equation}\label{hub}
\dot{Q}^{+} = 
\theta \; \frac{\dot{\boldsymbol{\eta}}}{\boldsymbol{\eta}} \cdot
[ \; \langle \; \upsilon(\cdot) \; \daleth \;
\rangle 
- \langle \; \upsilon(\cdot) \; \rangle
\langle \daleth \;
\rangle \; ] 
\end{equation}
where the angle brackets denote the average is taken over the set (\ref{questates}) with respect to
measure (\ref{queProb}), 
\begin{equation}
\daleth = \log \Big( ( \frac{\pi_{1}}{q_{1}})^{\frac{\alpha}{N} - \upsilon_{1}}
( \frac{\pi_{2}}{q_{2}})^{\frac{\alpha}{N} - \upsilon_{2}} \cdots 
( \frac{\pi_{M}}{q_{M}})^{\frac{\alpha}{N} - \upsilon_{M}}\Big)
\end{equation}

and

\begin{equation}
\upsilon(\cdot) = \Big\{  \upsilon_{1} ,\upsilon_{2} ,\ldots,\upsilon_{M} \Big\}
\end{equation}

denotes an element of the set $\boldsymbol{\upsilon}$ from (\ref{questates}).\footnote{
Of course, $
\langle \upsilon(\cdot)
\rangle
$
is simply $\Upsilon$ from (\ref{bigupsilon}).}

The derivation of (\ref{hub})
is of the same type as the derivation of
the work rate presented above except that the starting point
is (\ref{heatrate}) instead of (\ref{Workrate}).

The heating rate stands unadorned

\begin{equation}%\label{heatrate} 
\dot{Q} = \dot{p} \cdot \mathbf{H}
\end{equation}

where recall,

\begin{equation}%\label{H.} 
H_{k}=\theta \, \log \Big( \,\frac{\Pi}{ x_{k}} \,\Big),
\end{equation}

and

\begin{equation}%\label{intro_x} 
x_{k} = \frac{\pi_{k}}{\mathsf{q}_{k}},
\; k=1,2,\ldots,N.
\end{equation}

The $\pi$'s and $q$'s are obtained by observing
the trajectory on $\upsilon$. This definiton of heating
rate is consistent with the usual notion of heat as
a microscopic quantity.

Note that when the system is at equilibrium
(\ref{timeprob}) and (\ref{energyprob}) are directly
comparable and $\dot{Q}$ is the heating rate
per (\ref{dotp1s_equal}), (\ref{dotp2s_equal})
and (\ref{heatrate}).

\end{document}